\documentclass[preprint,english,showpacs,11pt,floatfix]{revtex4}

\usepackage{babel,amsmath,amssymb,dcolumn}
\usepackage[dvips]{graphics}
\usepackage{hyperref}

\usepackage{amsfonts}
\usepackage{amssymb}
\usepackage[dvips]{graphicx}
\usepackage{latexsym}

\usepackage{dcolumn}
\usepackage{bm}

\begin{document}

\title{Dark Energy: relating the evolution of the universe from the past to the future}
\author{Zhuo-Yi Huang, Bin Wang}
\email{wangb@fudan.edu.cn} \affiliation{Department of Physics,
Fudan University, Shanghai 200433, People's Republic of China }

\author{Rong-Gen Cai}
\email{cairg@itp.ac.cn} \affiliation{Institute of Theoretical
Physics, Chinese Academy of Sciences, \\
P.O. Box 2735, Beijing 100080, China}

\author{Ru-Keng Su}
\email{rksu@fudan.ac.cn} \affiliation{China Center of Advanced
Science and Technology (World Laboratory), P.B.Box 8730, Beijing
100080, People's Republic of China
\\Department of Physics, Fudan University, Shanghai 200433,
People's Republic of China }

\begin{abstract}

Using the evolution history of the universe, one can make
constraint on the parameter space of dynamic dark energy models.
We discuss two different parameterized dark energy models. Our
results further restrict the combined constraints obtained from
supernova and WMAP observations. From the allowed parameter space,
it is found that our universe will experience an eternal
acceleration. We also estimate the bound on the physically
relevant regions both in the re-inflationary and inflationary
phases.
\end{abstract}


\maketitle

Numerous and complementary cosmological observations made in the
last decade suggest that the expansion of our universe is now
accelerating (re-inflation)~\cite{01}. This may indicate that our
universe contains  dark energy with equation of state
$w_{de}<-\frac{1}{3}$, making up as much as 70\% of the critical
energy density. The usual suspects for dark energy are
cosmological constant with constant equation of state $w_{de}=-1$
or exotic fields with time dependent equation of state \cite{02}.

In addition to telling us contents of our universe, acceleration
of the cosmic expansion has also profound implications for
dynamics and physics on both very high energy scale in the past
and low energy scale in the future. Using the characteristic
length scale of expansion, the Hubble radius $H^{-1}$, together
with the energy density in terms of the expansion factor, we can
distinguish our universe into three different phases starting from
inflationary expansion (inflation), following by radiation and
matter dominated phase in standard big bang cosmology and finally
entering the re-inflationary era governed by the dark energy. For
an eternally accelerating third phase driven by the dark energy,
the event horizon surrounding any observer will limit the number
of e-folds of the re-inflation that will be observable in our
universe. More interestingly, it was found that when a universe is
at the moment of the transition to the eternally re-inflationary
phase, it contains the most information about the first phase
inflationary perturbations. This could impose a bound on the
physically relevant duration of the first phase inflation.
Considering the dark energy as the cosmological constant, the
concept of investigating the dynamic ranges during inflationary
and re-inflationary phases has been discussed in \cite{03,04}.
Besides, the phenomenology for the expansion dynamics in the
second phase of the universe evolution influenced by the dark
energy has also been examined in \cite{05}. It was argued that the
dark energy serves as a thread of three phases of the evolution of
our universe, relating the past to the future.

In this paper, we are going to generalize previous studies and
examine the dynamical implications on the past and future
expansion behaviors of our universe due to the dark energy with
time-dependent equation of state. The evolving dark energy is an
alternative model to the cosmological constant, which affects the
features of the temperature anisotropies in the cosmic microwave
background radiation and influences a lot on the small $l$ CMB
spectrum \cite{06,07}. In addition to its relevance to the early
universe evolution, we will show that the behavior of the
variation of equation of state for dark energy determines the fate
of our universe. Considering the evolution of energy density with
the expansion of the universe, we will present that the history of
the universe expansion on the contrary also puts constraints on
any possible variation of equation of state for dark energy. The
expansion of the universe and the evolution of dark energy are
closely co-related.

The expansion of the background universe is described by
\begin{equation}\label{eq1}
    H^2(a) = H_0^2 \left[ \Omega_r^0 a^{-4} + \Omega_m^0 a^{-3} + \Omega_d^0 f(a)
    \right],
\end{equation}
where $f(a)=\exp\left[ 3  \int_a^1
\frac{1+w_{de}(a')}{a'}\mathrm{d}a' \right]$, $\Omega_r$,
$\Omega_m$ are dimensionless radiation and matter densities and
$\Omega_d$ is the dark energy density. This Friedmann equation can
be used to give pictures of the second and third phases of the
evolution of the universe. For the first inflationary phase, we
simply assume that the Hubble parameter remains a constant
$H_{I}$. And  we further assume that this phase finished with a
perfectly efficient reheating so that the radiation era started
just at the end of the first phase with the scale factor
$a_{\mathrm{end}}$.

The effective total equation of state of the universe can be
calculated by \cite{05,08}
\begin{eqnarray}\label{eq2}
w_{tot}(a) & = & -1
-\frac{1}{3}\frac{\mathrm{d}\ln(H^2/H_0^2)}{\mathrm{d}\ln
a},\nonumber\\
 & = & \frac{\Omega_r^0 a^{-4} + 3\Omega_d^0 f(a) w_{de}(a)}
 {3\left[ \Omega_r^0 a^{-4} + \Omega_m^0 a^{-3} + \Omega_d^0 f(a)
 \right]}.
\end{eqnarray}
In the radiation era, $w_{tot} = \frac{1}{3}$. At present, since
we are already in the re-inflationary era according to
observations, $w_{tot}<-\frac{1}{3}$. The re-inflation started at
$z_* = 0.46 \pm 0.13$ at $1\sigma$ \cite{13}, which requires $q =
-\frac{a}{H}\frac{\mathrm{d}H}{\mathrm{d}a}-1$ crossing 0 and
puts $w_{tot} = -\frac{1}{3}$ at
$a_*=\frac{1}{1+z_*}$. All these historical moments of the
universe expansion could give constraints on the dark energy
evolution. We will show that combining the constraints from the
expansion history with supernova type Ia and WMAP observations, we
can severely restrict possible parameter space in the
time-dependent dark energy models.

In the far future, it is easy to see from Eq.~(\ref{eq2}) that
$w_{tot}\rightarrow w_{de}$. Thus the asymptotic behavior of the
evolving dark energy at large scale will determine the fate of the
evolution of our universe. The re-inflationary phase could keep on
forever or give way to the decelerated expansion provided that at
large scale the time-dependent equation of state approximates to
$w_{de}<-\frac{1}{3}$ or $w_{de}>-\frac{1}{3}$ respectively.

\begin{figure}[!hbtp] \label{fig1}
\begin{center}
\includegraphics[width=15cm]{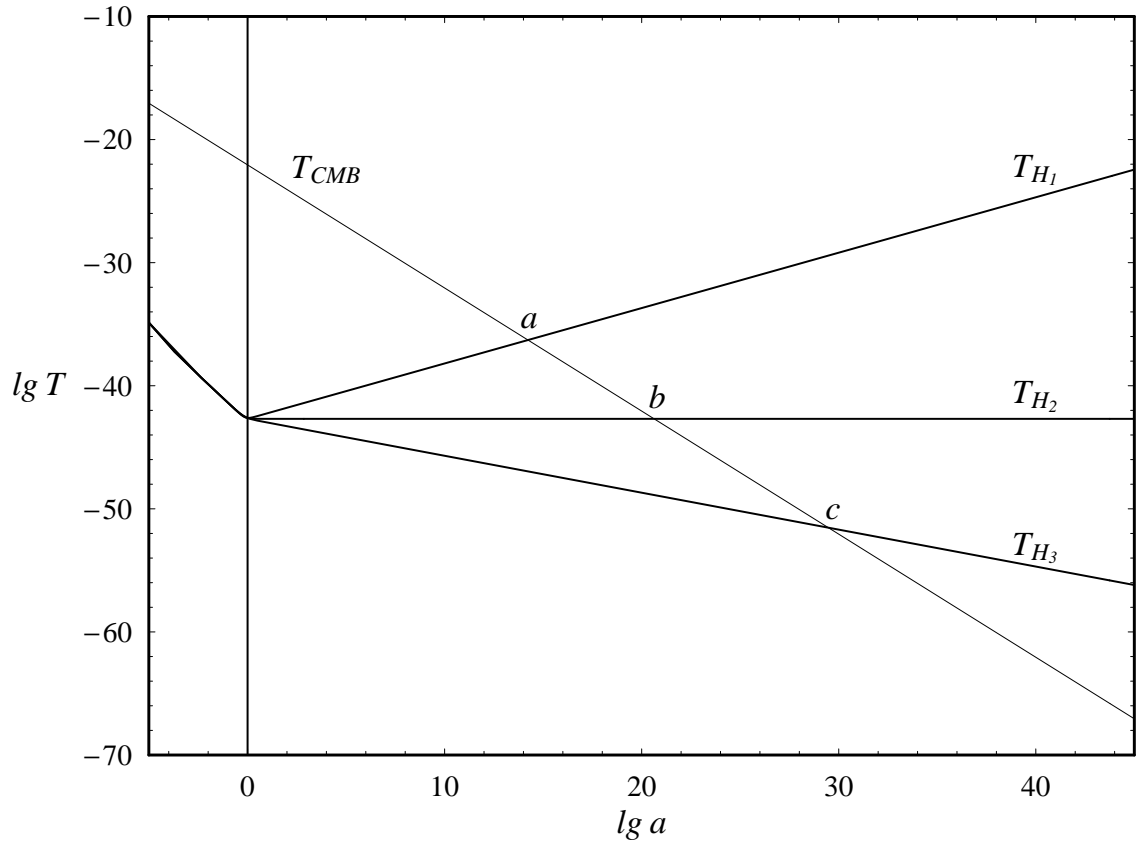}
\end{center}
\caption{Evolution of the temperature of the CMB $T_{CMB}$ and
Hawking particles $T_{H}$. Points $\mathit{a}$, $\mathit{b}$ and
$\mathit{c}$ represent the time that CMB information is immersed
in Hawking radiation for $w_{de}<-1$, $w_{de}=-1$ and $w_{de}>-1$
respectively .}
\end{figure}

If the universe ends with a decelerated expansion, a patient
observer could get as much information from CMB spectrum of our
universe as he/she observes. However, while the re-inflationary
phase lasts forever, the observer will always be surrounded by a
cosmological event horizon. Spacetimes with event horizons contain
Hawking particles. As the universe acceleratingly expands, the
wavelength of the CMB photons will  be redshifted rapidly and CMB
temperature will drop below the Hawking temperature. After this
the CMB information will be completely immersed in the
cosmological Hawking radiation. We illustrated this idea in Fig
1\ref{fig1}. Thus if the universe starts to accelerate forever in
the third phase, there will  be limited region of the universe
accessible to observers. For the eternally accelerating phase
driven by the cosmological constant, the investigation on the
cutoff scale in the observable region was carried out in
\cite{03,04}. Employing the combined constraints on the equation
of state of dark energy, we will estimate the finite physical
relevant region in the third phase for dynamic dark energy models.

Similar bounds on the physically relevant duration of the first
phase inflation also exist. It is known that the inflationary
perturbations could act as seeds of structure formation after they
re-enter the Hubble horizon. The wavelength of the quantum
fluctuations stretches as $\lambda \sim a(t)$, while after the
first phase, the Hubble horizon of the universe grows linearly in
time, which is much faster than the stretch of the wavelength
scale of the fluctuations. Hence normally the perturbations will
reenter the Hubble radius sooner or later after their eviction
from it if there was no re-inflation. If the universe enters the
third phase with forever acceleration, the Hubble radius will be
flattered (or shrunk, depending on the equation of state for dark
energy), grows even slower than the perturbation wavelength. Some
of the perturbations will never reenter the Hubble horizon. The
later the fluctuations reenter the Hubble radius, the 
earlier they exit from the Hubble horizon during the first phase.
The last moment for the fluctuations generated during inflation to
reenter the Hubble horizon as the universe enters the third phase
is exactly the moment when the re-inflation starts. Perturbations
crossing the Hubble horizon much earlier during inflation will
never reenter the Hubble horizon in the re-inflation phase.
Employing the criterion that we need the perturbation to reenter
the Hubble horizon, we can get the astrophysical relevant bound on
the length scale produced during the inflation. Clear picture of
this idea is shown in Fig 2 \cite{03}. Considering the third phase
of the universe evolution
 being driven by the dynamic dark energy, we will
find that the starting moment of the re-inflation will be shifted
from that of assuming the third phase being de-Sitter phase
\cite{03}. The shift of the starting point of the re-inflation
will change the bound on the duration of inflation obtained in
\cite{03}. This will be discussed in some detail for different
dynamical dark energy models below.

\begin{figure}[!hbtp] \label{fig3}
\begin{center}
\includegraphics[width=15cm]{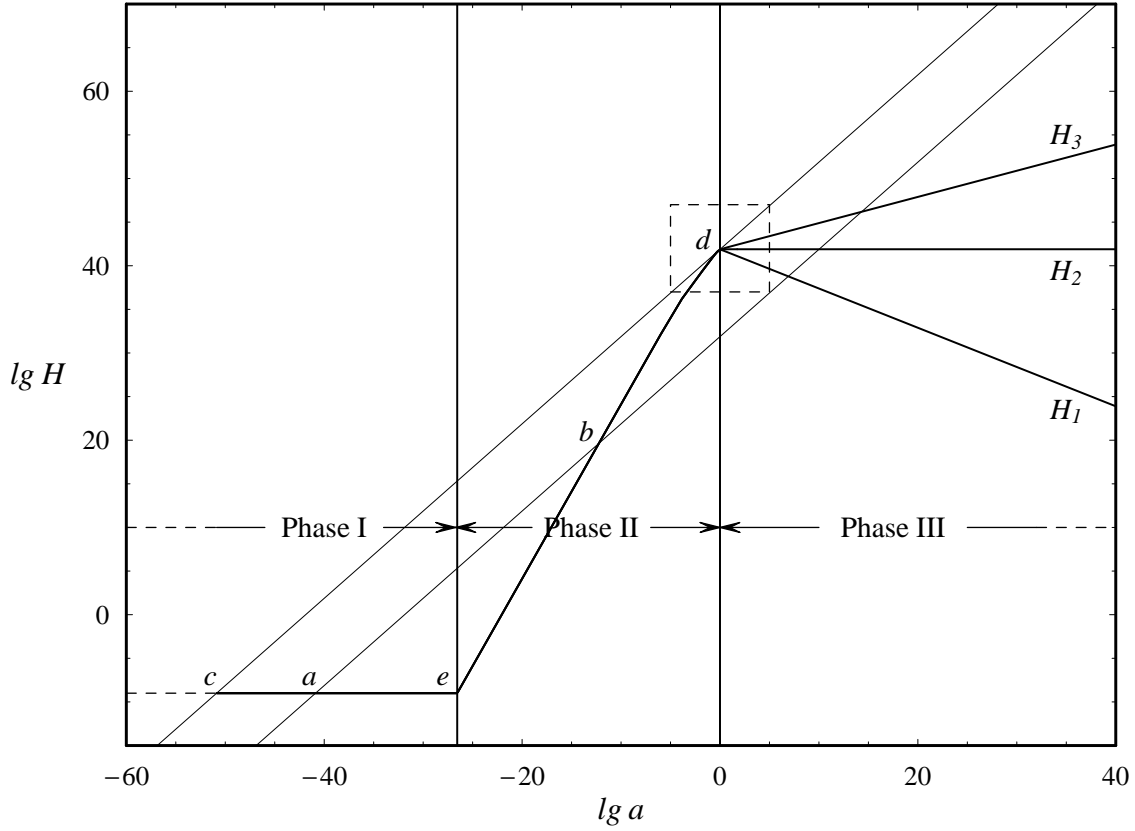}
\end{center}
\caption{The evolution of our universe experiences three phases,
namely inflation (line $ce$), radiation and matter domination
(line $ed$) and re-inflation. Point $a$ and $b$ show the time
perturbations generated during inflation exit and re-enter the
Hubble radius. The perturbations which leave Hubble radius earlier
than $c$ will never re-enter due to the acceleration in the third
phase. During re-inflation, $H_1$, $H_2$ and $H_3$ represent the
evolution of Hubble radius in the future with $w_{de}<-1$,
$w_{de}=-1$ and $w_{de}>-1$ respectively. The beginning moments of reacceleration 
are different due to the variations in the equation
of state for dark energy.}
\end{figure}

In our following discussion, we will choose two time-dependent
dark energy models with different parametrizations:
\begin{eqnarray}\label{eq3}
    w_{de}^{I}(a) & = & w_0 + w_1(1-a),\\
    w_{de}^{II}(a) & = & w_0 + w_1(1-a)a.
\end{eqnarray}
These two models have been extensively discussed in various
papers, for example, see \cite{06,09,10,11,12}. For the first
 model, characteristic moments of the universe
evolution require that during the radiation era
$w_{tot}=\frac{1}{3}$ and at the present time
$w_{tot}<-\frac{1}{3}$ due to the observed accelerated expansion,
which gives constraints: $w_1 \leq 0.34 - w_0$ and $w_0<-0.48$.
These constraints shrink the allowed parameter space of $w_0-w_1$
obtained from supernova observation. Using the observational
result that the re-inflation started at
$z_*=0.46\pm0.13$\cite{13}, which is the point where the
deceleration parameter $q =
-\frac{a}{H}\frac{\mathrm{d}H}{\mathrm{d}a}-1$ crossing 0, we can
have further constraints on the $w_0-w_1$ parameter space for this
dynamic dark energy model. The combined constraints from the
evolution history and the supernova and WMAP observations are
shown in the white region in Fig 3. The result shows that adding
the evolution history constraint, the variation of the equation of
state of dark energy can be severely restricted.

\begin{figure}[!hbtp] \label{fig2}
\begin{center}
\includegraphics[width=10cm]{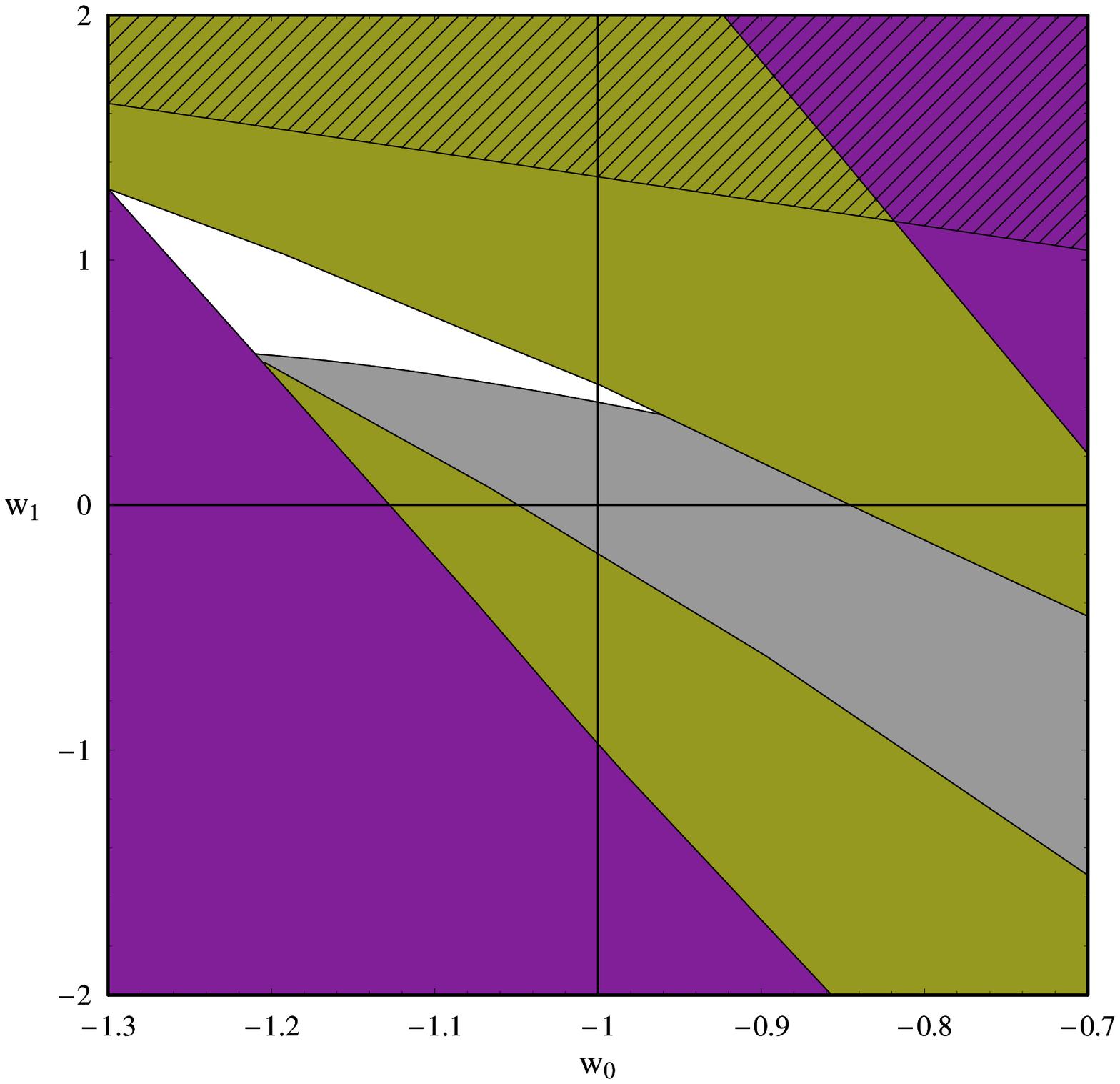}
\end{center}
\caption{The purple region is excluded by supernova observations
and the green part is ruled out by the WMAP observations
\cite{03}. The shaded region violets $w_{tot}=\frac{1}{3}$ at
radiation era. The gray part is dropped by requiring the
transition to the re-inflation phase happened at $z_*=0.46\pm 0.13$.}
\end{figure}

In the far future, we know that $w_{tot}\rightarrow w_{de}$ from
Eq.~(\ref{eq2}). The fate of the universe is determined by the
value of $w_1$. For $w_1<0$, the re-inflationary expansion of the
universe will give way to deceleration at very late time. However,
if $w_1>0$, $w_{tot}$ will  be a very negative value at large
scale, thus the re-inflation of the universe will  last forever.
The combined constraints shown in Fig 3 tell us that the allowed
$w_1$ is indeed positive leading to the eternal re-inflation of
the universe in the third phase of its evolution. Using the
concept explained above, the bound on the region of the universe
could be accessible to observers in the future is within the range
with e-folds $N_\mathrm{re-inf}\in [3.68, 7.91]$ by employing the
allowed parameter space in Fig 3.

On the other hand, by fixing the starting moment of the
re-inflation within the constraint parameter space of the evolving
dark energy and using the criterion that we need the fluctuations
to reenter the Hubble radius,  the natural bound on the duration
of the inflation with the number of e-folds is found to be
$N_{\mathrm{inf}}\in [55.98, 56.02]$.

These studies can be extended to the second parametrization of the
dynamic dark energy model. The universe evolution history requires
(\textbf{a}) at the radiation era $w_{tot}=\frac{1}{3}$,
(\textbf{b}) the deceleration parameter crosses $0$ at $a_*$,
(\textbf{c}) at present  $w_{tot}<-\frac{1}{3}$. Especially the
requirement (b) can be used to further restrict the constraints
got by the supernova and WMAP observations. The combined
constraint is shown in the white region in Fig 4 \ref{fig3}.

\begin{figure}[!hbtp] \label{fig4}
\begin{center}
\includegraphics[width=10cm]{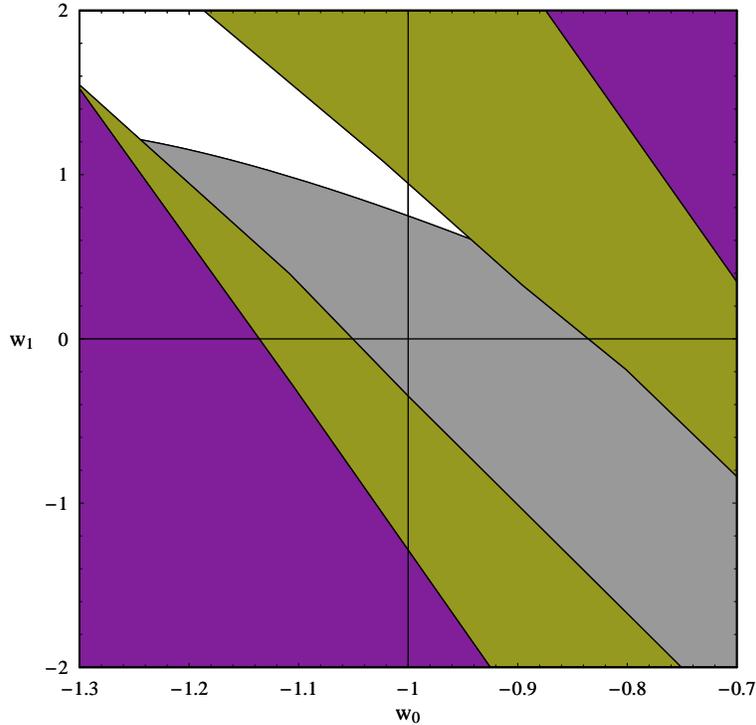}
\end{center}
\caption{Similarly as explained in Fig 3, the white region left is
the combined constraints resulting from both the history of the
universe expansion and supernova and WMAP observations.}
\end{figure}

Similar to the first model, the sign of $w_1$ determines the fate
of the universe. For $w_1<0$, the present re-inflation era will be
followed by a deceleration expansion. However for $w_1>0$, the
present accelerated expansion will go on forever. From the
combined constraint shown in Fig 4 \ref{fig3}, we learn that our
universe will experience an eternal acceleration. The bound on the
number of e-folds  of the future universe, which is observable, is
found within the range $N_{\mathrm{re-inf}}\in [2.29, 2.87]$. In
addition, using the criterion discussed above, the natural cutoff
length scale in the inflation is estimated to be within the number
of e-folds $N_{\mathrm{inf}}\in [55.95, 56.01]$. This number is
quite near to that in the first model.

In summary, we investigated the close relation between the
expansion history of the universe and the evolution of the dynamic
dark energy. We found that some specific historical moments of the
universe evolution can be used to constrain  the parameter space
of the evolving dark energy models, which  further refines the
combined constraints from supernova and WMAP observations. The
behavior of the evolving dark energy can determine the fate of the
universe.  For two dynamic dark energy models with different
parametrizations, we found that the universe will experience an
eternal acceleration. In addition, the natural cutoff length
scales in both the inflation and re-inflation phases were
obtained, which make regions of physical relevance in both of
these two phases finite.

\begin{acknowledgments}
This work was partially supported by  NNSF of China, Ministry of
Education of China and Shanghai Education Commission.
\end{acknowledgments}


\end{document}